\documentclass[11pt]{amsart}

\begin{document}

\title[]{The Total Space-Time of a Point Charge and Its Consequences for
Black Holes}

\author{Leonard S. Abrams}

\address{24345 Crestlawn Street, Woodland Hills, California 91367}

\email{ag272@lafn.org}
\thanks{Published in Int. J. Theor. Phys. {\bf 35} (1996) pp. 2661-2677}

\begin{abstract}
Singularities associated with an incomplete space-time $S$
are not uniquely defined until a boundary $B$ is attached to it.
[The resulting space-time-with-boundary, $\bar{S}$ $\equiv S\cup
B$, will be termed a ``total'' space-time (TST).] Since an incomplete
space-time is compatible with a variety of boundaries, it follows
that S does not represent a unique universe, but instead
corresponds to a family of universes, one for each of the distinct
TSTs. It is shown here that the boundary attached to the
Reissner-Nordstr\"om space-time for a point charge is invalid for
$q^2<m^2$. When the correct boundary is used, the resulting
TST is inextendible. This implies that the Graves-Brill black hole
cannot be produced by gravitational collapse. The same is true of
the Kruskal-Fronsdal black hole for the point mass, and for those
black holes which reduce to the latter for special values of their
parameters.
\end{abstract}

\maketitle

\section{INTRODUCTION}

As is well known \cite{Tipler80}, the singularity structure of a
casually incomplete, not necessarily maximal space-time $S=
(M,g)$ is not uniquely specified by $(M,g)$ alone, but requires in
addition that a boundary $B$ be attached
thereto \cite{HE73}.\footnote[1]{For simplicity, whenever an order
of differentiability is required to completely specify a concept
(e.g., manifold, extension, equivalent, etc), ``analytic'' is always
to be understood.} [The resulting object, $T\equiv S\cup B$, will
be termed a `total' space-time (TST).] Since an incomplete
space-time is compatible with a variety of boundaries
(Geroch, 1968, p. 451) \cite{Geroch68}, it follows that $S$
by itself cannot represent  a unique universe - instead, it
corresponds to a family of universes,
one for each of its topologically distinct boundaries.\par
Recognition of this fact immediately gives rise to the question:
"Which of the possible boundaries of $S$ should be attached to it?".
The answer depends on how much additional information is
available; if nothing more is known than $(M,g)$ itself, then one is
free to attach any of its mathematically admissible boundaries, the
only remaining problem being the physical interpretation of the
result. If, on the other hand, one also knows the universe $U$ from
which $S$ was derived, or which the TST of $S$ is intended to
represent, then there is no freedom whatever to choose $B$ - the
boundary is uniquely determined by $U$, just as are such things as
the symmetries of $S$, whether $S$ is static, etc. In some cases this
determination could be quite difficult, while in others it could be
trivial. An example of the latter situation would be a $U$
consisting of a single point source, for then the boundary of the
TST of $S$ is evidently a line through the location of the
source.\par Incidentally, note that since a TST is a
space-time-with-boundary, the criteria for equivalence and
extendibility of TST are necessarily different from those
applicable to space-times. Specifically, equivalence of TSTs
requires not only that their $S$ components be isometric
\cite{SW77}\footnote[2]{As can be seen from p. 26 of this
reference, equivalence also requires that the isometry be
orientation- and time-orientation-preserving; hence any reference
herein to an isometry will be understood to mean one having these
properties.}, but also that their $B$ components be homeomorphic.
Likewise, extendibility of a TST requires not only extendibility of
its $S$ component, but also that its $B$ component be preserved - i.e.,
that the image of $B$ be homeomorphic to $B$ itself.\par
Admittedly, in most cases the only difference between the
TSTs associated with different choices of a boundary
for a given $S$ is the topology of
the singularities, as is the case in the example in Hawking and
Ellis \cite{HE73}. However, when $S$ is extendible, the choice of
boundary to be attached to it may well affect the very existence
of a singularity of the resulting TST. For example, consider the
following two-dimensional, Riemannian case: $(M,g)=(D_{1},g_{E})$,
where $D_{1}$ denotes the interior of the unit disk
in $R^{2}$ (i.e., $x^{2}+y^{2} <  1$), and $g_{E}$ denotes the
Euclidean metric: $ dx^{2}+ dy^{2}$. If the boundary is taken to
be $S^1$, then there is no singularity at the boundary and the
space possesses an extension to all of $R^{2}$, namely $(R^{2},
g_{E})$. On the other hand, if the boundary is taken to be a point
(e.g., by identifying all the points of the circle $x^{2} + y^{2}
= 1)$, then there is a quasiregular singularity at the boundary
and no extension is possible.\par As is well known, the
Reissner-Nordstr\"om space-time $S_{RN}$ is timelike
incomplete \cite{Carter63}, so that by the argument in the first
paragraph it cannot represent a unique universe. Moreover,
examination of Nordstr\"om's derivation shows that it involves a
tacit assumption which is invalid for $q^2<m^2$ and which attached
a boundary to $ S_{RN}$ that is incompatible with the point-charge
universe from which $S_{RN}$ was derived. As a result, the RN TST
does not represent a point charge for $q^2<m^2$, and {\it a
fortiori}, neither does its black-hole-containing
extension \cite{GB60}. When the space-time for a point charge
is correctly derived and the appropriate boundary is attached, the
resulting TST is inextendible and devoid of a black hole. Since
this TST is necessarily the limit of the space-time of any
spherically symmetric, nonrotating, charged star that is
collapsing to a point, it follows that such collapse cannot produce
a black hole.\par Since there is no way of deriving the
Graves-Brill (GB) space-time from the set of postulates
characterizing a point charge, nor of producing it by
gravitational collapse, it follows that it has neither a
theoretical nor a physical justification for its existence. Thus
it, and the black hole which it contains, are simply artifacts of
a historical error. The same is true of the point-mass black
hole \cite{Fronsdal59}, \cite{Kruskal60}, since it is based on
Hilbert's space-time for a point mass, whose derivation (in 1917)
involved the same erroneous
assumption \cite{Abrams89}\footnote[3]{In the referenced paper, I
asserted (\S~6) that since Schwarzschild's and Hilbert's
space-times had different singularity structures, it followed that
the two space-times were inequivalent (cf. Sachs and
Wu \cite{SW77}, \S~1.3). This is incorrect - as shown here, it
is only their TSTs which are inequivalent.} as was used by
Nordstr\"om.\par The purpose of this paper is to derive the
correct TST for a point charge and to show how the above-mentioned
consequences come about.\par The plan of the paper is as follows:
Section 2 is devoted to showing that Nordstr\"om's
\cite{Nordstrom18} derivation made use of an unjustified
assumption, which among
other things resulted in the attachment of a boundary that is
incompatible with the point-charge universe when $q^2<m^2$.
Section 3 lists the historical postulates on which Reissner's
\cite{Reissner16} and Nordstr\"om's derivations were based.
Section 4 contains a derivation of the point-charge
space-time for $q^2<m^2$
which does not make use of Nordstr\"om's assumption, and in the
process proves that the assumption is invalid for the case in
question. In addition, it also contains a proof that the
historical postulates do not give rise to a unique space-time, but
to a one-parameter family of inequivalent space-times, and thus
that these postulates must be supplemented by one which determines
the limiting value of a certain invariant as the point charge is
approached. Section 5 discusses the physical significance of the
supplemental hypothesis, and proposes a criterion for choosing
among the infinitely many possible values of the invariant. In
Section 6 the metric which results from this choice is obtained,
and the associated TST is shown to be equivalent to that found by
Pekeris \cite{Pekeris82}. Section 7 deals with the
phenomenon of a nonspinning, spherically symmetric, charged star
undergoing gravitational collapse to a point. It is shown that
such collapse can never produce a GB black hole. Section 8 shows
that because of the infinite red shift at the horizon of the GB
space-time, there are in principle no phenomena explainable by a
GB black hole that would not be equally well explained by Pekeris'
TST, so that the former is unnecessary from a phenomenological
standpoint (and considerably more complicated than Pekeris' TST).
Finally, the principal conclusions of the paper are summarized in
Section 9, together with a brief discussion.

\section{HISTORICAL BACKGROUND}
The first attempt to determine the space-time of a point
charge was that of Reissner \cite{Reissner16}, However, his
derivation was only valid for the case $q^2>m^2$, where $m$
denotes the particle's field-producing mass (i.e., that which
produces Kepler-like orbits far from the point charge) and $q$
denotes its charge (in relativistic esu) In addition, it contained
an unknown constant [$h(0^+)$ in Reissner's notation] whose value
was tacitly assumed.\par
A year or so later, a derivation applicable to all values of
${q^2}/{m^2}$ was presented by
Nordstr\"om \cite{Nordstrom18}. Taking the point charge to
be at $x=y=z=0$, the starting point of his derivation was the
static, spherically symmetric metric
\begin{equation}\label{2.1}
g_{sss}(r|0) \equiv A(r)dt^2-B(r)dr^2-C(r)d\Omega^2
\end{equation}
where
\begin{equation}\label{2.2}
r \equiv (x^2+y^2+z^2)^{1/2}
\end{equation}
and
\begin{equation}\label{2.3}
d\Omega^2 \equiv d\theta^2 + d\phi^2\sin ^2{\theta }
\end{equation}
\par[The $r|0$ in the argument of $g_{sss}$ signifies that in
terms of $r$, the position of the point charge is described by $
r=0$. A similar notation will be used for subsequent metrics.]\par
Substituting equation (\ref{2.1}) into the variational equivalent
of the Einstein-Max\-well field equations gave Nordstr\"om three ordinary
differential equations (ODE) for the three unknown coefficients.
Instead of solving these (which is done here in Appendix A),
Nordstr\"om argued that one could always introduce a new radial
variable, $r^*$ say, via
\begin{equation}\label{2.4}
r^*=[C(r)]^{1/2}
\end{equation}
This transformation carries equation (\ref{2.1}) into
\begin{equation}\label{2.5}
g^*_{sss}(r^*|
r^*_0)=A^*(r^*)dt^2-B^*(r^*)dr^{*2}-r^{*2}d\Omega^2,
\end{equation}
and, as follows directly from equation (\ref{2.4}), assigns the value
\begin{equation}\label{2.6}
r^* = r^*_0 = [C(0+)]^{1/2}
\end{equation}
to the location of the point charge. Setting $C^*=r^{*2}$ into
the above-mentioned ODE, Nordstr\"om then solved for $A^*$ and
$B^*$, obtaining expressions for these quantities which contained
but one constant of integration, $\alpha_N$. This constant was
readily shown to be equal to $2m$, giving the results
\begin{equation}\label{2.7}
A^*(r^*) = A_N(r^*) \equiv 1 - 2m/r^* + q^2/r^{*2}
\end{equation}
\begin{equation}\label{2.8}
B^*(r^*) = B_N(r^*) \equiv 1/A_N(r^*)
\end{equation}
\par
However, the fact that $C(r)$ is unknown at the stage where $r^*$
is introduced shows that the value of $r^*_0$ cannot be determined
from equation (\ref{2.4}). Moreover, the fact that $r = 0$ is the location
of the point charge and thus a singularity of the field makes it
impossible to appeal to `elementary flatness' to conclude that
$C(0+)=0$. Consequently, Nordstr\"om's solution seemingly
involves an unknown and unknowable constant, namely the $r^*$
coordinate of the point charge. [This would appear, for example,
in the expression for the proper distance from the point charge to
an event having $r^* = r_P^*$, namely
\begin{equation}\label{2.9}
d(r_P^*) = \int^{r_P^*}_{r_0^*}[B^*(r^*)]^{1/2}dr^*
\end{equation}
or in that for the proper volume from the point charge to $r_P^*$,
etc.] However Nordstr\"om {\it tacitly} assumed that\footnote[4]{In
\cite{Nordstrom18}, just under equation (5), he states, ``..where
$e(r^*)$ denotes the total charge in a sphere of radius $r^*$.''
And just above equation (19a), he states, ``Because of the
spherical symmetry we have, of course for the components of $U$ in
the directions of the axes of coordinates in space
\begin{eqnarray}\nonumber
U_\tau = x_{\tau}c{\alpha}/({r^*}^3) &\tau = 1,2,3''
\end{eqnarray}
having previously defined $U = c{\alpha}/{(r^{*2})}$ in equation
(19). [{\it Note:} $r$ in  Nordstr\"om's paper has been changed to
$r^*$ here so as to conform to the notation adopted in the present
paper.] Neither of these statements makes sense unless $r^*$ is
regarded as ${(x^2+y^2+z^2)}^{1/2}.$} $r^*={(x^2+y^2+z^2)}^{1/2}$,
which implies that the location $(x=y=z=0)$ of the point charge in
terms of $r^*$ is given by $r^*=0$. But for $q^2<m^2$, $B_N^*$ tends
to $\infty$ as $r^* \downarrow r^*_+ \equiv m+ {(m^2 - q^2)}^{1/2}$,
so for this case
Nordstr\"om's metric is only defined for $r^* > r^*_+$. Moreover,
this assumption also results {\it in the attachment of a
particular boundary} to this space-time: a 2-sphere at $r^* =
r^*_+$ in the sections $t =$ const.\par Although Nordstr\"om only
claimed that his space-time represented a point charge for
$q^2>m^2$, this was lost sight of in the ensuing years, and
$S_{RN}$ has come to be regarded as the space-time of a point
charge for all values of ${q^2}/{m^2}$. Following historical
usage, the metric obtained from equation (5) by setting $A^*=A_N,$
$B^*=B_N$, {\it and} assuming that $ r^*={(x^2+y^2+z^2)}^{1/2}$
will henceforth be termed the Reissner-Nordstr\"om
metric and will be denoted by $g_{RN}(r^*|0)$:
\begin{equation}\label{2.10}
g_{RN}(r^*|0) = A_N(r^*)dt^2 - B_N(r^*)dr^{*2} -r^{*2}d\Omega^2
\end{equation}
\par
Subsequent derivations of the point-charge metric were carried out
by H\"onl and Papapetrou \cite{Honl39} and Pekeris
\cite{Pekeris82}. In both cases the authors applied the
tecnique used by Schwarzschild \cite{Schwarz16} in his
derivation of the point-mass metric, thereby avoiding
Nordstr\"om's error; but in each case {\it ad hoc} assumptions
were used to evaluate one of the integration constants and thereby
obtain a unique field.
\par
Analytic extensions of the RN space-time were subsequently
obtained by Graves and Brill \cite{GB60} for
${q^2}<{m^2}$ and by Carter \cite{Carter63} for
${q^2}={m^2}$.\par Finally, the point of view adopted here is the
generally accepted one that charge of either sign increases the
field-producing mass, and that charge without matter is
impossible; the first of these implies that $q^2\leq m^2$ ,
while the second reduces the latter to $q^2<m^2$. (This
implies that point-charge space-times for which $q^2\geq m^2$
are nonexistent, and thus so, too, is the Carter black hole.) This
inequality is to be understood to hold in the following sections.

\section{THE HISTORICAL POSTULATES}
The postulates that until now have been regarded as characterizing
the point-charge space-time $S_{PC} \equiv (M_{PC},g_{PC})$ may be
gleaned from Reissner's paper. In vernacular terms, they require
that:\par
(i) $M_{PC}$ consist of $R^4$ less a line through the point
charge ($\equiv M_0$).\par
\medskip
That $g_{PC}$ be:\par
\medskip
(ii) Of Lorentz signature.\par
(iii) Analytic.\par
(iv) Static.\par
(v) Spherically symmetric about the point charge.\par
(vi) A solution of the Einstein-Maxwell field equations.\par
(vii) Flat at spatial infinity (i.e., $g_{ij}\rightarrow
\eta_{ij}$).\par
\medskip
And that:\par
\medskip
(viii) $S_{PC}$ be inextendible.

\section{THE EXTENT TO WHICH THE HISTORICAL POSTULATES DETERMINE
THE METRIC}

Let $K$ denote $R^4$ considered as the analytic manifold arising
from the sin\-gle-chart atlas ($R^4,Id$), and ($t,x,y,z$) the
natural coordinates (O'Neill, 1983, p. 1) \cite{O'Neill83} thereon.
Without loss of generality we shall take the charge to be located at
$x=y=z=0$. Let $L$ denote the line $x=y=z=0$, so that
$M_0=K\backslash L$.\par
As is well known \cite{Eiesland25}, a metric satisfying (ii)-(v)
everywhere on $M_0$ can always be expressed in the form
\begin{eqnarray}\nonumber
g_{PC}(x,y,z|0,0,0) = A(r)dt^2 -F(r)(dx^2 +dy^2 +
dz^2)\\\label{4.1}
-G(r){(xdx +ydy +zdz)}^2
\end{eqnarray}
where
\begin{equation}\label{4.2}
r={(x^2+y^2+z^2)}^{1/2}
\end{equation}
and $A$, $F$, $G$ are analytic functions of $r$, satisfying
\begin{equation}\label{4.3}
A,F,G>0
\end{equation}
so as to ensure compliance with (ii).\par
Transforming to quasipolar coordinates related to $x,y,z$ in the
customary way, we find that equation (\ref{4.1}) becomes
(with a slight abuse of notation)
\begin{equation}\label{4.4}
g_{PC}(r|0) = A(r)dt^2-B(r)dr^2-C(r)d\Omega^2
\end{equation}
while condition (\ref{4.3}) becomes
\begin{equation}\label{4.5}
A,B,C>0
\end{equation}
Substituting equation (\ref{4.4}) into the Einstein-Maxwell
equations, solving for $A$, $B$, and $C$, and imposing (vii) gives
(see Appendix A)
\begin{eqnarray}\label{4.6}
A(r)=1-2m/{C^{1/2}}+q^2/C
\end{eqnarray}
\begin{eqnarray}\label{4.7}
B(r)={C'}^2/{(4AC)}
\end{eqnarray}
where the prime denotes differentiation with respect to $r$. These
expressions coincide with those found by Stavroulakis \cite{Stav81}
for the exterior of a static, spherically symmetric charged body
when the cosmological constant in his result is set to zero, and
with $q=0$, to those obtained for the point mass in
Oliver \cite{Oliv77} (this paper was the first in which the
point-mass field equations were integrated without eliminating one
of the unknowns).\par
However, as shown in Appendix A, for a point
charge, $C$ must also be an analytic function of $r$ such that
\begin{eqnarray}\label{4.8}
C(r)>0 &{\rm for} &r>0
\end{eqnarray}
\begin{eqnarray}\label{4.9}
C'(r)>0 &{\rm for} &r>0
\end{eqnarray}
\begin{eqnarray}\label{4.10}
C/{r^2}\rightarrow 1 &{\rm as} &r\rightarrow\infty
\end{eqnarray}
\par
All that is necessary in order to comply with the remaining
historical postulates is to impose  whatever further conditions
are required to ensure that $A$ is positive [it is also necessary
that $B$ be positive on $M_0$, but conditions
(\ref{4.7})-(\ref{4.9}) show that this is assured if $A>0$ for $r>0$]
and that (viii) is satisfied.\par
As to the first of these, Appendix B shows that $A$ will be positive
iff
\begin{equation}\label{4.11}
\sqrt{C}>r^*_+\equiv m+(m^2-q^2)^{1/2}
\end{equation}
and since this must hold for all $r>0$, it follows that
\begin{equation}\label{4.12}
b\equiv{[C(0+)]}^{1/2}\geq r^*_+>0
\end{equation}

Since this condition is necessary to ensure the positivity of $A$,
{\it it follows that Nordstr\"om's assumption, which requires
that $b=0$, is invalid}.\par As to the second, the fact that the
metric is analytic everywhere on $M_0\equiv K\backslash L$
shows that (viii) will
be satisfied only if $L$ is singular. Now, Appendix C shows that
the Kretschmann scalar $f=R_{ijkm}R^{ijkm}$ tends to infinity as
$C\rightarrow 0$ and is bounded otherwise. In view of
equations (\ref{4.9}) and (\ref{4.12}), this means that there are no
scalar curvature singularities on $L$. But as is well known, it is
possible to have a singularity without any curvature-related
scalar being unbounded [a so-called quasiregular singularity (Ellis
and Schmidt, 1977, p. 944) \cite{ES77}]. To explore this
possibility, let us first observe that the boundary $r=0$ must be
a point in each spatial section, so as to comply with the fact
that the associated TST is intended to represent a universe
consisting of a single point source. Next, suppose that $b\geq
r^*_+$, and consider the geodesic circle $\gamma:t=t_0,~
\theta=\pi/2,~ r=\epsilon$. Its proper circumference clearly tends
to $2{\pi}b>0$ [from equation (\ref{4.4})] as
$\epsilon\downarrow 0$, while its proper radius is easily seen to
tend to zero, so that the ratio of the former to the latter tends
to $\infty$. This shows that $r=0$ is a singularity, since such
behavior makes it impossible to map an infinitesimal neighborhood
of $r=0$ into an interior portion of some larger space-time.
(Note that the image of $r=0$ under any such map must be a point
in each spatial section, since otherwise the boundary attached to
$M_0$ would not be homeomorphic to its image in the map, as is
required for extendibility of a TST.) Consequently, postulate
(viii) is satisfied for any value of $b\geq r^*_+$. [Despite
appearances, the foregoing argument is coordinate independent,
since equation (\ref{4.4}) is unique up to transformations of the
form $t=k\bar{t}+p$, $r=h(\bar{r})$, and neither of these
alters the proper radius or proper circumference of $\gamma$.] It
is the presence of this singularity which makes the TST
inextendible.\par
Moreover, just as in the case of the point
mass \cite{Abrams89}, so here, too, the value of $C[P]$ at any
event $P$ is a scalar invariant of $g_{PC}(r|0)$, and thus the
same is true of its limit $(b^2$) as $P$ approaches the point
charge. Consequently, space-times whose metrics are of the form of
$S_{PC}$ with distinct values of $b$ are inequivalent. Hence we
conclude:
\begin{itemize}
\item The historical postulates do not lead to a unique
space-time, but to a one-parameter family of inextendible,
inequivalent space-times. In order to obtain a unique field, it is
necessary to supplement those postulates by one which fixes the
value of $C(0+)$.
\end{itemize}

\section{THE SUPPLEMENTAL POSTULATE}
We have seen in the previous section that in order to arrive at a
unique field, it is necessary to supplement the historical
postulates for the point charge by one which fixes the value of
$b={[C(0+)]}^{1/2}$. Now, as shown in Appendix D, the limiting
acceleration of a neutral test particle approaching the point
charge on a radial geodesic of equation (\ref{4.4}), as measured
by a sequence of fixed observers, is given by
\begin{equation}\label{5.1}
a_0={\frac{|{bm-q^2}|}{{b^2}\sqrt{b^2 -2{bm}+q^2}}}\equiv
{\frac{|{bm-q^2}|}{{b^2}\sqrt{{(b-m)}^2+q^2-m^2}}}
\end{equation}
    From this it is evident that fixing the limiting value of the
locally measured acceleration of a radially approaching neutral
test particle is equivalent to fixing $b$. For the same reason as
stated in connection with the point mass \cite{Abrams89}, we shall
take this limiting acceleration to be the same ($\infty$) as in
the Newtonian case, thereby supplementing the historical
postulates with the following:\par (ix) The limiting value of the
locally measured acceleration of a neutral test particle
approaching the point charge on a radial geodesic is
infinite.
\section{THE RESULTING METRIC}
Clearly, the only value of $b$ consistent with equation
(\ref{4.12}) which makes the RHS of equation (\ref{5.1}) infinite
is
\begin{equation}\label{6.1}
b=r^*_+=m+({m^2-q^2})^{1/2}
\end{equation}
\par The simplest $C$ satisfying the earlier requirements as well
as equation (\ref{6.1}) is given by
\begin{equation}\label{6.2}
C_P\equiv( {r+r^*_+})^2
\end{equation}
which reduces equations (\ref{4.6}) and (\ref{4.7}) to
respectively
\begin{eqnarray}
&A_P(r)=1-2m/{(r+r^*_+)}+q^2/{{(r+r^*_+)}^2}\\\label{6.3}
&B_P(r)={[A_P(r)]}^{-1}\label{6.4}
\end{eqnarray}
It is readily verified via the transformation
$r+r^*_+={{[\bar{r}}^3+{(r^*_+)}^3]}^{1/3}$ that the resulting
space-time is isometric to that of Pekeris, and since the boundary
atttached to the latter is likewise pointlike in any spatial
section, the two TSTs are equivalent. Moreover, since Nordstr\"om's
and the just-derived version of Pekeris' space-times are isometric
via the transformation $r^*=r+r^*_+$, it follows that the
{\it actual} location of the point charge in terms of $r^*$ is
given by $r^*=r^*_+$, so that the $2$-sphere
boundary resulting from Nordstr\"om's assumption {\it coincides
with the position of the point charge}, which is clearly
incompatible with the pointlike nature of the source.

\section{GRAVITATIONAL COLLAPSE}
Eiesland \cite{Eiesland25} showed that any spherically symmetric
solution of the Ein\-stein-Maxwell field equations is static. In
particular, then, this last must be true of the `exterior' metric
of a charged, nonrotating, nonradiating spherically symmetric
object (`star') undergoing gravitational collapse. Consequently,
this metric must satisfy the same postulates as for the point
charge, namely (viii) and (ix). As can be seen from Appendix A,
satisfaction of postulates (ii)-(vii) when applied to a metric of
the form of equation (\ref{4.4}) entails that $A$ and $B$ be given
by equations (\ref{4.6}) and (\ref{4.7}), respectively, and
together with condition (\ref{4.5}), that $C$ be a
positive, analytic, strictly monotonic increasing function of $r$
over the $r$ range for which the exterior metric is valid -~i.e.,
for $r>r_b(t)$, where $r_b(t)$ denotes the $r$ coordinate of the
star's boundary at time $t$.\par
Since the set of $C$'s which,
together with equations (\ref{4.6}) and(\ref{4.7}), give rise to a
metric satisfying (ii)-(vii) for $r>r_b(t)$ may obviously be
larger than that satisfying (ii)-(ix) for $r>0$, it cannot be
asserted that the exterior metric is the `same' as that of the
point charge (in the sense of having the same values of $C$ at the
same proper distances from the center of symmetry). However,
if the star is sufficiently massive as to collapse to a point,
then as it does so, the difference between the actual exterior
metric and the metric of the point charge must vanish as
$r\downarrow 0$. Consequently, in that limit the exterior metric
becomes that of the point charge. Since the point-charge space-time
whose metric was obtained in the previous section has no event
horizon, it follows that a GB black hole cannot be formed during
the collapse of a charged star. Thus, the correction of
Nordstr\"om's error not only eliminates the point charge as a
possible source of such black holes, but simultaneously deprives
the latter of the only mechanism that has been proposed for their
production.

\section{THE GB BLACK HOLE IS UNNECESSARY}
Just as in the case for the point mass, so here, too, there is an
infinite red shift in the GB space-time as $r^*\downarrow r^*_+$.
Because of this, all that an outside observer can ever know of
phenomena involving the GB black hole must arise from information
originating outside the hole - i.e., from the RN space-time. But
the latter is diffeomorphic to that of Pekeris - that is to say,
everything that takes place outside the hole would occur in the
identical fashion if the GB space-time were replaced by Pekeris';
it is impossible to determine which space-time is ``really'' present
without going to the boundary. Thus, any observations that could be
explained by postulating the presence of a GB black hole would be
equally explained by postulating the existence of a Pekeris `point'
charge at the boundary of the black hole. Consequently, there is
no need to invoke a GB black hole to explain any set of
observations. Pekeris' black `point' will do an equally effective
job, and its topology is far simpler.

\section{CONCLUSIONS}
1. Nordstr\"om's assumptions that the radial coordinate in his
metric is related to $x$, $y$, $z$ via $r^*=
(x^2+y^2+z^2)^{1/2}$ is not only invalid, but also leads to the
attachment of a boundary to his space-time that is incompatible
with the pointlike nature of the source.\par
2. The historical postulates restrict the manifestly static and spherically
symmetric form of the point-charge metric to equation (\ref{4.4})
cum equations (\ref{4.6})-(\ref{4.10}), where (with $b\equiv
{[C(0+)]}^{1/2}$)
\begin{equation*}
b\geq r^*_+=m+({m^2-q^2})^{1/2}
\end{equation*}
Thus, those postulates give rise to a one-parameter family of
inequivalent space-times; to obtain a unique field, they must be
supplemented by one which fixes the value of $b$. Since the minimum
value of $b$ is necessarily $\geq r^*_+>0$ in order to ensure the
Lorentz signature of the metric, it follows that {\it however this
supplemental postulate is chosen}, Nordstr\"om's assumption, which
requires that $b=0$, is invalid.\par
3. The limiting value $a_0$ of
the locally measured acceleration of a neutral test particle
approaching the point charge on a radial geodesic fixes $b$. The
supplemental postulate therefore requires a choice of $a_0$, which
on the basis of the Newtonian analogy is here taken to be
infinite.\par
4. The point-charge metric resulting from this choice is
\begin{eqnarray}\nonumber
g_P&=[1-2m/{(r+r^*_+)}+q^2/{{(r+r^*_+)}^2}]dt^2\\\nonumber
&-[1-2m/{(r+r^*_+)}+q^2/{{(r+r^*_+)}^2]^{-1}}dr^2\\\nonumber
&-{{(r+r^*_+)}^2d\Omega}^2 &{\rm for}~r>0
\end{eqnarray}
whose associated TST is equivalent to Pekeris'.\par
5. The `exterior' metric for a charged, spherically symmetric star
undergoing gravitational collapse is of the same form as that for
a point charge, and as the collapse proceeds must approach $g_P$.
Thus no black hole can be formed by such collapse.\par
6. Any observations that would be consistent with the existence of
a GB black hole would be equally consistent with a Pekeris black
`point' and thus GB black holes are unnecessary to explain any
phenomena.\par
7. Possessing neither a theoretical justification
(i.e., a mathematical chain {\it from} a set of postulates
characterizing a particular universe {\it to} the GB black hole)
nor a mechanism for its production, the GB black hole is nothing
more than an artifact of a historical error.\par
8. Conclusion 7 applies equally well to the Kruskal-Fronsdal black hole,
since by setting $q=0$ in  Nordstr\"om's TST reduces it to that
erroneously derived by Hilbert for the point mass \cite{Abrams89} and
reduces the GB space-time to that obtained by Kruskal \cite{Kruskal60}
and Fronsdal \cite{Fronsdal59}. It also applies \cite{Abrams96} to the
black hole found by Lake and Roeder \cite{LR77} for the point
mass when $\Lambda\neq 0$.\par
It should be noted that what has been termed Pekeris' black `point'
does not have all the properties of a point in Euclidean 3-space
- in particular, although the proper radius and proper volume of a
sphere about $r=0$ passing through $r=\epsilon$ go to zero as
$ \epsilon \downarrow 0$, the proper circumference and proper area of
the sphere approach nonzero values. Were it to have {\it all} the
properties of an ordinary point, of course, it would not be a
singularity of the TST, as is required of a point source.\par
For $q=0,$ the TST obtained here was also found by Janis {\it et al.}
\cite{JNW68} on the basis of entirely different considerations.

\numberwithin{equation}{section}
\appendix
\section{THE MOST GENERAL SOLUTION SATISFYING
THE HISTORICAL POSTULATES}

Making the obvious modifications in the equations
[Tolman, (1934) p. 259, equation (102.6)] \cite{Tolman34} satisfied
by the electromagnetic field tensor $F_{ij}$ in empty space,
necessitated by the use of $C$ rather than $r^2$ in equation
(\ref{2.1}), we find as usual that all $F^{ij}$
vanish except for $F^{41}=-F^{14}$, and that the former satisfies
(a prime denotes differentiation with respect to $r$)
\begin{equation}\label{A.1}
(\ln F^{41})'+{[\ln({C\sqrt{AB}})]}'=0
\end{equation}
whence
\begin{equation}\label{A.2}
\phi'=-ABF^{41}=-K{\sqrt{AB}}/C
\end{equation}
where $\phi$ denotes the `time' component of the generalized
potential and $K$ is an integration constant. Substituting these
results into the expression [Tolman, p. 261, equation (104.1)]
\cite{Tolman34} for Maxwell's energy-momentum tensor, we obtain
\begin{equation}\label{A.3}
T^1_1=-T^2_2=-T^3_3=T^4_4=K^2/{(2C^2)}
\end{equation}
whence the nonvanishing and independent field equations read (cf.
\cite{Abrams89})
\begin{equation}\label{A.4}
-8\pi T^1_1=-1/C+C^{'2}/(4BC^2)+{A'C'}/(2ABC)=-f_E
\end{equation}
\begin{eqnarray}\label{A.5}
-8\pi T^2_2=&C''/(2BC)+A''/(2AB)-{C'}^2/(4BC^2)\\\nonumber
&-B'C'/(4{B^2}C)-A^{'2}/(4{A^2}B)-A'B'/(4AB^2)\\\nonumber
&+A'C'/(4ABC)=f_E\nonumber
\end{eqnarray}
\begin{eqnarray}\label{A.6}
-8\pi T^4_4&=C''/(BC)-1/C-B'C'/(2{B^2}C)-{C'}^2/(4BC^2)\\\nonumber
&=-f_E\nonumber
\end{eqnarray}
where
\begin{equation}\label{A.7}
f_E\equiv 4\pi K^2/{C^2}
\end{equation}
\par
Subtracting equation (\ref{A.6}) from equation (\ref{A.4}) leads,
as in the uncharged case, to
\begin{equation}\label{A.8}
C^{'2}=JABC
\end{equation}
where $J$ is an integration constant. Substituting this into
equation (\ref{A.4}) gives
\begin{equation}\label{A.9}
{A'+{\frac{C'}{2C}}A}=\bigg(\frac{1}{C}-\frac{4\pi
K^2}{C^2}\bigg)\frac{2C'}{J}
\end{equation}
which integrates at once to
\begin{equation}\label{A.10}
A=4/J-\alpha/{\sqrt C} +{16\pi K^2}/(JC)
\end{equation}
with $\alpha$ being the constant of integration. Finally, using
equation (\ref{A.8}) to eliminate $B$ from equation (\ref{A.5})
and eliminating $A$ from the result via equation (\ref{A.10})
shows, after some algebra, that equation (\ref{A.4}) is satisfied
identically. Thus, equations (\ref{A.8}) and
(\ref{A.10}) constitute the general solution of equations
(\ref{A.4})-(\ref{A.7}).\par
Now, application of postulate (vii) to equation (\ref{A.1}) shows
that $C$ must behave like $r^2$ as $r\rightarrow \infty$.
Hence, from equation (\ref{A.10}), $A\rightarrow 4/J$ as
$r\rightarrow \infty$ and thus, by (vii) again, $J=4$.
Moreover, substituting equation (\ref{A.8}) into the
RHS of equation (\ref{A.2}) and letting $r\rightarrow \infty$
gives
\begin{eqnarray}\label{A.11}
\phi'\sim-K/r^2 &{\rm as} &r\rightarrow \infty
\end{eqnarray}
whence
\begin{equation}\label{A.12}
K=q/(4\pi)^{1/2}
\end{equation}
where $q$ denotes the charge on the particle in relativistic esu
($c=G=1$).\par

Collecting these results, we have
\begin{equation}\label{A.13}
A=1-2m/{\sqrt C}+q^2/C
\end{equation}
\begin{equation}\label{A.14}
B={C'}^2/{(4AC)}
\end{equation}
where $\alpha$ has been identified with twice the charge's
field-producing mass by comparison with the Newtonian potential
for large $r$. From equations (\ref{A.14}) and (\ref{A.4}) we see
that $C'$ cannot be zero for $r>0$, so that it must always be
positive or always be negative. Of these two possibilities the
latter is ruled out by the requirement that $C\sim r^2$ for large
$r$. Consequently we conclude:
\begin{itemize}
\item $C$ is any positive, analytic, strictly monotonic increasing
function of $r$ that behaves like $r^2$ as $r\rightarrow\infty$.
\end{itemize}
The only other constraints on $C$ that are a consequence of the
historical postulates are those induced by the necessity of
satisfying equation (\ref{4.5}) and (viii). These matters are
dealt with in Appendix B.

\section{ADDITIONAL CONSTRAINTS ON {\it C} DUE TO THE
HISTORICAL POSTULATES}

As shown in Appendix A, $C$ must be positive and strictly
monotonic-increasing. Hence by equation (\ref{A.14}), $B$ will be
positive iff $A$ is. Thus the satisfaction of condition (\ref{4.5})
requires only that $C$ be chosen so as to make $A>0$. Rewriting
equation (\ref{A.13}) in the form
\begin{equation}\label{B.1}
A=\frac{(\sqrt{C}-m)^2+q^2-m^2}{C}
\end{equation}
we see that in order to make the numerator of equation (\ref{B.1})
positive and thus comply with condition (\ref{4.5}), it is
necessary that either
\begin{equation}\label{B.2}
\sqrt{C}>m+(m^2-q^2)^{1/2}\equiv r^*_+
\end{equation}
or
\begin{equation}\label{B.3}
\sqrt{C}<m-(m^2-q^2)^{1/2}\equiv r^*_-
\end{equation}
But, since it was shown in Appendix A that $C$ must behave like
$r^2$ as $r\rightarrow\infty$, it follows that $C$ cannot satisfy
equation (\ref{B.3}) for all $r>0$. Consequently, the satisfaction
of (\ref{4.5}) imposes the additional requirement of inequality
(\ref{B.2}) on $C$. Finally, inspection of equation (\ref{B.1})
shows that for $C$ constrained by inequality (\ref{B.2}), the
Kretschmann scalar $f=R_{ijkm}R^{ijkm}$ is well behaved as
$r\rightarrow 0$, and thus (since the spherical symmetry ensures
that all other scalars are necessarily functions of $f$) there is
no curvature-related singularity there.

\section{EVALUATION OF $f$}
The nonvanishing components of $R_{ijkm}$ for equations
(\ref{2.1})-(\ref{2.3}) are readily found to be
\begin{eqnarray}\label{C.1}
R_{1212}=(2BCC''-B'CC'-B{C'}^2)/(4BC)\\
R_{1313}=R_{1212}sin^2\theta\\
R_{1414}=(B{A'}^2+AA'B'-2ABA'')/(4AB)\\
R_{2323}=({C'}^2-4BC)sin^2\theta /(4B)\\
R_{2424}=-A'C'/(4B)\\
R_{3434}=R_{2424}sin^2\theta
\end{eqnarray}
together with their counterparts obtained by interchange of the
indices in the first pair or the second pair.\par
The corresponding values of $R^{ijkm}$ are

\begin{eqnarray}\label{C.7}
R^{1212}=R_{1212}/(B^2C^2)\\
R^{1313}=R_{1313}/(B^2C^2sin^4\theta)\\
R^{1414}=R_{1414}/(A^2B^2)\\
R^{2323}=R_{2323}/(C^4sin^4\theta)\\
R^{2424}=R_{2424}/(A^2C^2)\\
R^{3434}=R_{3434}/(A^2C^2sin^4\theta)
\end{eqnarray}
and their associated counterparts.\par
Taking account of the multiplicity of terms obtainable by
interchanging indices, it readily follows that
\begin{eqnarray}\nonumber
f=4(R_{1212}R^{1212}+R_{1313}R^{1313}+R_{1414}R^{1414}\\
+R_{2323}R^{2323}+R_{2424}R^{2424}+R_{3434}R^{3434}\\
=f_1+f_2+f_3+f_4\label{C.14}
\end{eqnarray}
where
\begin{eqnarray}
f_1\equiv 2[C''-B'C'/(2B)-{C'}^2/(2C)]^2/(B^2C^2)\\
f_2\equiv[{A'}^2/(2A)+(A'B')/2B-A'']^2/(A^2B^2)\\
f_3\equiv4[{C'}^2/(4B)-C]^2/C^4\\
f_4\equiv {A'}^2{C'}^2/(2A^2B^2C^2)
\end{eqnarray}
Elimination of $B$ via equation (\ref{A.14}) reduces these to
\begin{eqnarray}
f_1=8{A'}^2/{C'}^2\\
f_2=16{A'}^2C^2[A''/A'-C''/C'+C'/(2C)]^2/{C'}^4\\
f_3=4(A-1)^2/C^2\\
f_4=f_1
\end{eqnarray}
Finally, elimination of $A$ via equation (\ref{A.13}) gives
\begin{eqnarray}
f_1=f_4=8(m/C^{3/2}-q^2/C^2)^2\\
f_2=4(3q^2/C^2-2m/C^{3/2})^2\\
f_3=4(q^2/C-2m/{\sqrt C})^2/C^2
\end{eqnarray}
which upon substitution into equation (\ref{C.14}) give
\begin{equation}
f=8[6(m{\sqrt C}-q^2)^2+q^4]/C^4
\end{equation}

\section{EVALUATION OF $a_0$}

As shown in Doughty (1981) \cite{Doughty81}, the acceleration of a
neutral test particle relative to a fixed observer in a universe
whose metric can be written in the form of equation (\ref{4.4}) is
given by
\begin{equation}
a={\frac{\sqrt{-g_{rr}}(-g^{rr})|g_{tt,r}|}{2g_{tt}}}
\equiv\frac{|A'|}{2A\sqrt B}
\end{equation}
Using equation (\ref{A.13}) to obtain $A'$ and equation
(\ref{A.14}) to eliminate $B$, this gives
\begin{equation}
a=\frac{|m{\sqrt C}-q^2|}{C\sqrt{AC}}
\end{equation}
As $r\rightarrow 0$, this approaches
\begin{equation}
a_0={\frac{|{bm-q^2}|}{{b^2}\sqrt{b^2 -2{bm}+q^2}}}
\end{equation}
\vbox to 1.0 cm{}
\*{ACKNOWLEDGMENTS}\par
\vbox to 0.1 cm{}
It is a pleasure to acknowledge several exchanges of
correspondence with G.F.R. Ellis and R. Geroch, as well as
numerous conversations with B. O'Neill and R. Greene. Such errors
as remain are, of course, mine alone.

\end{document}